\documentclass[12pt]{iopart}
\usepackage{geometry}              
\usepackage{graphicx}
\usepackage{amssymb}
\usepackage[square, comma, sort&compress, numbers]{natbib}
\usepackage[font={small}]{caption}
\usepackage{epstopdf}
\begin{document}

\title[Hierarchy in Gene Expression is Predictive for Adult Acute Myeloid Leukemia]
{Hierarchy in Gene Expression is Predictive of Risk, Progression, and Outcome in
Adult Acute Myeloid Leukemia}
\author{Shubham Tripathi\textsuperscript{1} and Michael W. Deem\textsuperscript{2,3}\\
\textsuperscript{1}Department of Biological Sciences and Bioengineering\\
Indian Institute of Technology, Kanpur, UP 208016, India\\
\textsuperscript{2}Department of Bioengineering and
\textsuperscript{3}Department of Physics \& Astronomy\\
Rice University,
Houston, TX  77005, USA}

\begin{abstract}

Cancer progresses with a change in the structure of the gene network in normal cells. We define a measure of organizational hierarchy in gene networks of affected cells in adult acute myeloid leukemia (AML) patients. 
With a retrospective cohort analysis based on the gene expression profiles of 116 acute myeloid leukemia patients,
we find that the likelihood of future cancer relapse and the level of clinical risk are directly correlated with the level of organization in the cancer related gene network. We also explore the variation of the level of organization in the gene network with cancer progression. We find that this variation is non-monotonic, which implies the fitness landscape in the evolution of AML cancer cells is nontrivial. We further find that the hierarchy in gene expression at the time of diagnosis may be a useful biomarker in AML prognosis.

\end{abstract}


\bibliographystyle{iopart-num}

\submitto{\PB}
\maketitle

\section{Introduction}

Cancer cells break the most basic rules of cell behavior by which multicellular organisms are built and maintained, and they exploit every kind of opportunity to do so \cite{Albert}. They proliferate without restraint and prosper at the expense of their neighbors in a multicellular environment. The development of cancer typically requires that a substantial number of independent, rare genetic and epigenetic accidents occur in the lineage of one cell. Cancer, therefore, involves changes in the pattern of gene expression in normal cells. It involves disruption in the normal mechanisms of gene control and regulation, which results in correlated gene networks in a cell going astray. Since cancer progresses with changes in the gene network, the structure of the gene network in cancer cells may be predictive of clinical risk and outcome \cite{Taylor2009,Chuang2007,Pavlidis2002,Doniger2003,Draghici2003,Subramanian2005,Tian2005,Wei2007,Rapaport2007}.

Acute myeloid leukemia (AML) is the most common acute leukemia in adults. Chemotherapy induces a complete remission in 70 to 80 percent of patients aged 16 to 60 years, but many of them have a relapse and die of their disease. Myeloablative conditioning followed by allogeneic stem-cell transplantation can prevent relapse, but this approach is associated with a high treatment-related mortality. Therefore, accurate predictors of the clinical outcome are needed to determine appropriate treatment for individual patients \cite{Bullinger2004,Giles2002}. Currently used prognostic indicators include age, cytogenetic findings, the white-cell count, and the presence or absence of an antecedent hematologic disorder \cite{Lowenberg}.

Cancer is a microevolutionary process. We are motivated to study the hierarchical structure in cancer affected cells by a theory that relates development of hierarchy during evolution to environmental stress and variability \cite{Sun2007,Deem,Lorenz2011,Deem2013}. According to this theory, during evolution in a rugged fitness landscape amidst a changing environment a system tends to become more hierarchical, since hierarchy tends to increase the adaptability of the system. Therefore, we expect more aggressive cancer cells to exhibit a higher level of hierarchical organization in their cancer related gene network. Cancers with a more hierarchical cancer related gene network are expected to be more adaptable and competitive, and thus have a higher chance of relapsing.
We here provide support for this theory in the case of AML.

Structure in gene networks in cancer cells has been quantified, and it has been found to be useful in cancer prognosis. Vineetha \emph{et al.}\ applied a TSK-type recurrent neural fuzzy approach to extract regulatory relationship among genes and to reconstruct gene regulatory network from microarray data from colon cancer cells. Structure in this regulatory network provided new insights into colon cancer diagnostics \cite{Vineetha}. Taylor \emph{et al.}\ used the co-expression of hub proteins and their partners to identify whether interactions are context-specific, or constitutive. They found that loss of co-regulation in cancer results in disruptions in hub protein components of interaction networks \cite{Taylor2009,Chuang2007}. Ford \emph{et al.}\ used gene expression microarray data to predict the recurrence time in lung cancer \cite{Ford}.

Hierarchy is a measure of modularity that exists in gene expression networks at different levels. Here, we quantify the hierarchy in cancer related gene networks and explore its relation with clinical risk and outcome in AML.
That is, we study the relation between cancer relapse and the structural features of the gene network in cancer affected cells. 
We will show that the level of organization and hierarchy in the cancer related gene network in AML affected cells can serve as a biomarker in AML prognosis.
There is a significant difference in the hierarchy measure between patients that have a cancer relapse (34 patients) and those that do not (43 patients) ($p$-value=0.0012). The area under the ROC curve for the classifier that discriminates between the relapse and no relapse patient cohorts on the basis of the hierarchy measure is $0.70$, which is significant compared to a random classifier.

\section{Methods}

\subsection{Patients}
We performed a retrospective cohort analysis based on the gene expression profiles of acute myeloid leukemia patients (samples provided by the AML Study Group, Ulm, Germany). The gene expression data were previously obtained from samples obtained from 116 AML patients (median age, 50 years) \cite{Bullinger2004}. 
Of the various genes analyzed, 133 genes predictive of clinical outcome were identified \cite{Bullinger2004}. Relapse in patients was monitored during the follow up visits.
Subsequent reports, working with other datasets,  have also
reported gene expression data in AML associated with prognosis
\cite{Radmacher,Metzeler,Li}.

During the follow up period, 68 of the 116 patients died, and 34 of the 79 patients who had complete remission relapsed. 20 patients were also reported to have developed a refractory cancer disease. For the patients who died during the follow up period, the overall survival since diagnosis was also reported.

\subsection{Cytogenetics}
Cytogenetic analysis can be used to classify AML patients as being at low risk [t(8;21), t(15;17), or inv(16)], intermediate risk [e.g., a normal karyotype or t(9;11)], or high risk [e.g., inv(3), -5/del(5q), -7, or a complex karyotype (three or more aberrations)] \cite{Grimwade1998,Byrd2002,Slovak2000}. We categorized the patients into these categories depending on the cytogenetic aberration reported for each patient \cite{Bullinger2004}.

\subsection{Gene networks for different patient categories}

To study the structure and organization of the gene network in different patient groups, we defined a network consisting of 133 genes previously identified as being predictive of clinical outcome in AML \cite{Bullinger2004} as nodes. The weight of the link between any two nodes was defined using the Pearson correlation coefficient as
\begin{equation}
l_{\alpha,\beta}=\sum_i^n\frac{(P_{i,\alpha}-\mu_{\alpha})(P_{i,\beta}-\mu_{\beta})}{\sigma_{\alpha}\sigma_{\beta}}
\label{1}
\end{equation} 
Here $P_{i,\alpha}$ is the expression data of gene $\alpha$ for patient $i$ from  \cite{Bullinger2004}, $\mu_{\alpha}$ is the average expression value for gene $\alpha$ for the $n$ patients, and $\sigma_{\alpha}$ is the standard deviation of expression value of gene $\alpha$ for the $n$ patients. 
Using the above definition, we make comparisons between the following sets of categories of patients:

\subsubsection{Relapse, remission with no relapse, and refractory disease groups}
\begin{figure}[tbh!]  
\begin{center}  
a)\includegraphics[scale=0.35]{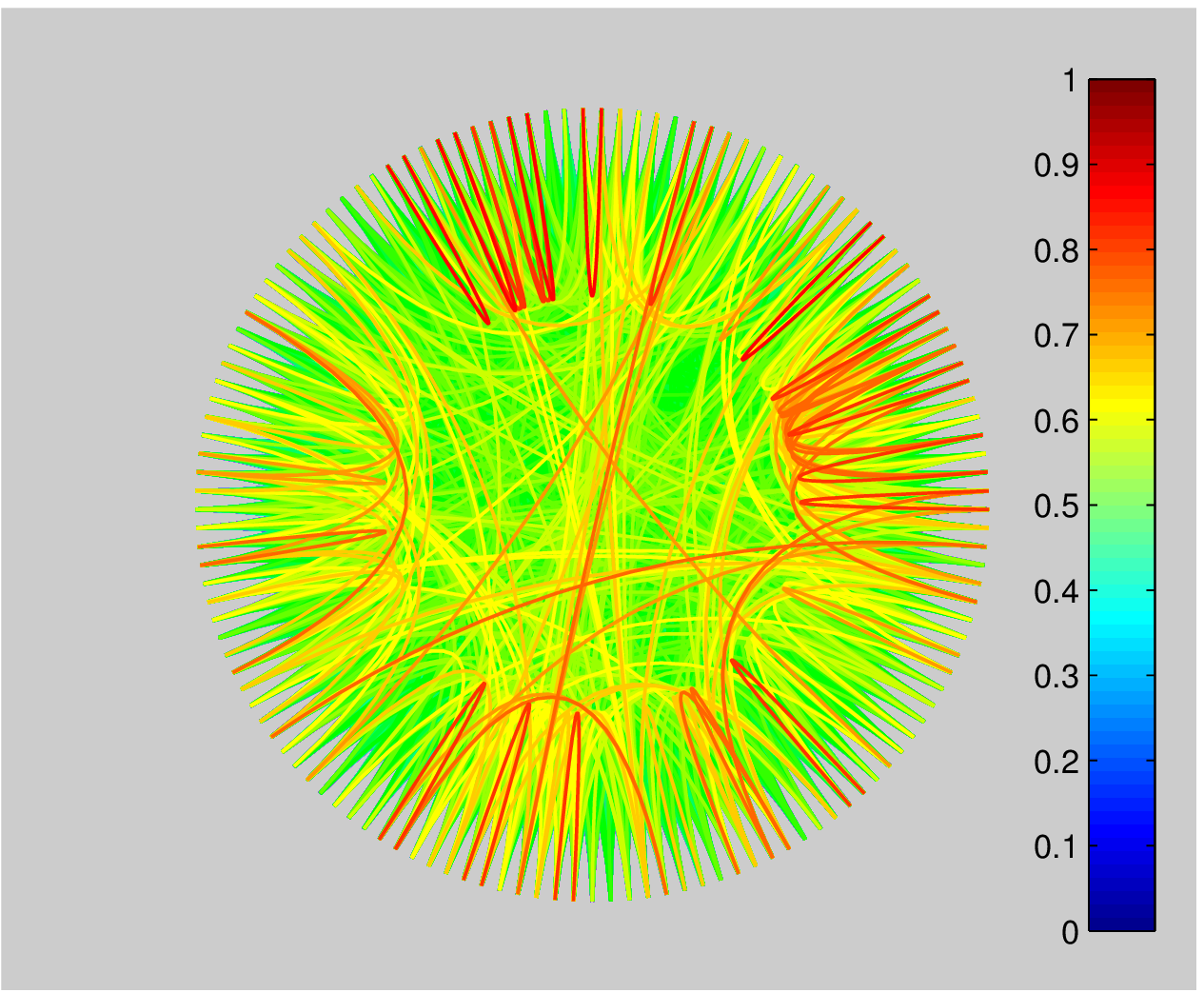}  
b)\includegraphics[scale=0.35]{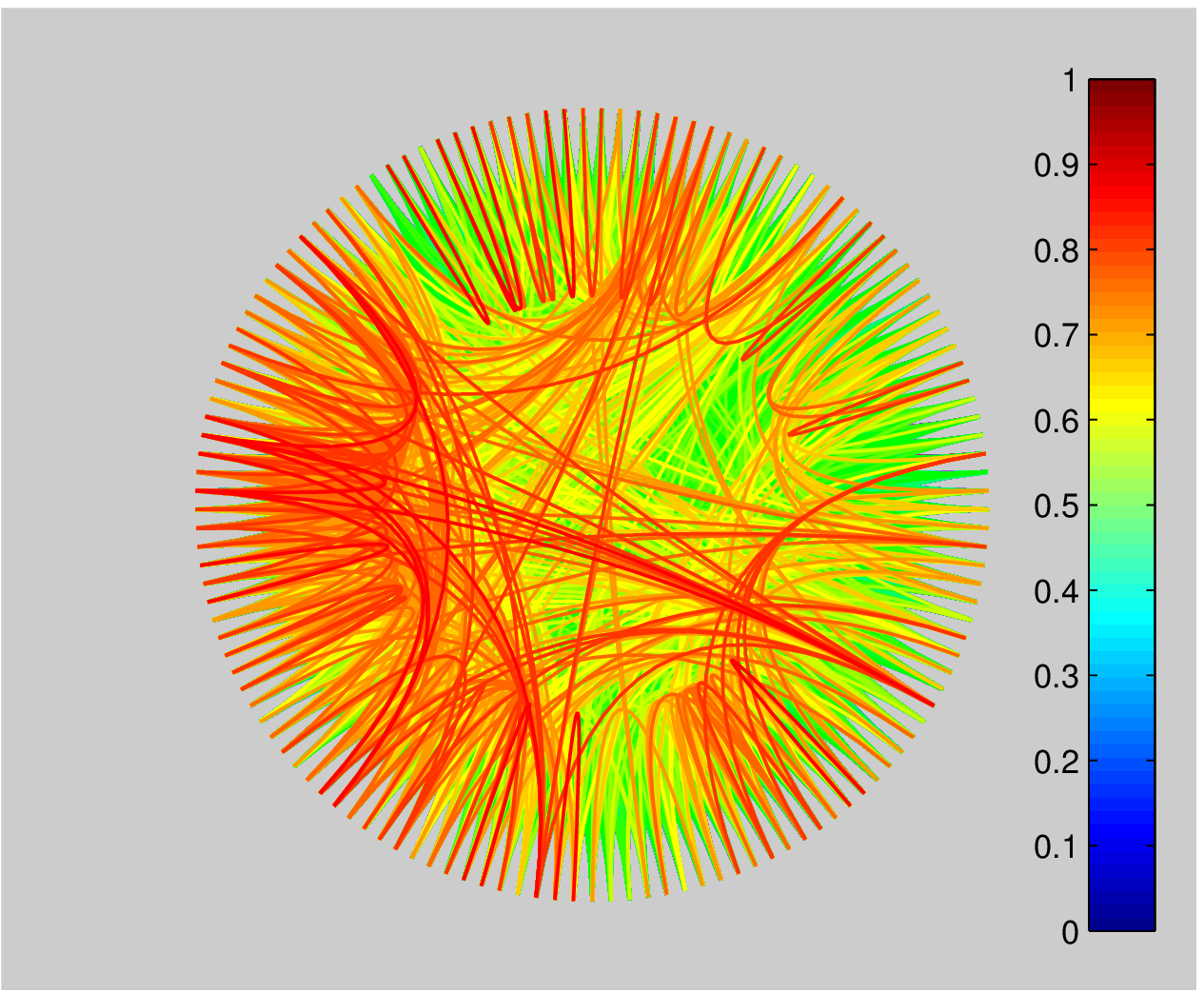}  
c)\includegraphics[scale=0.35]{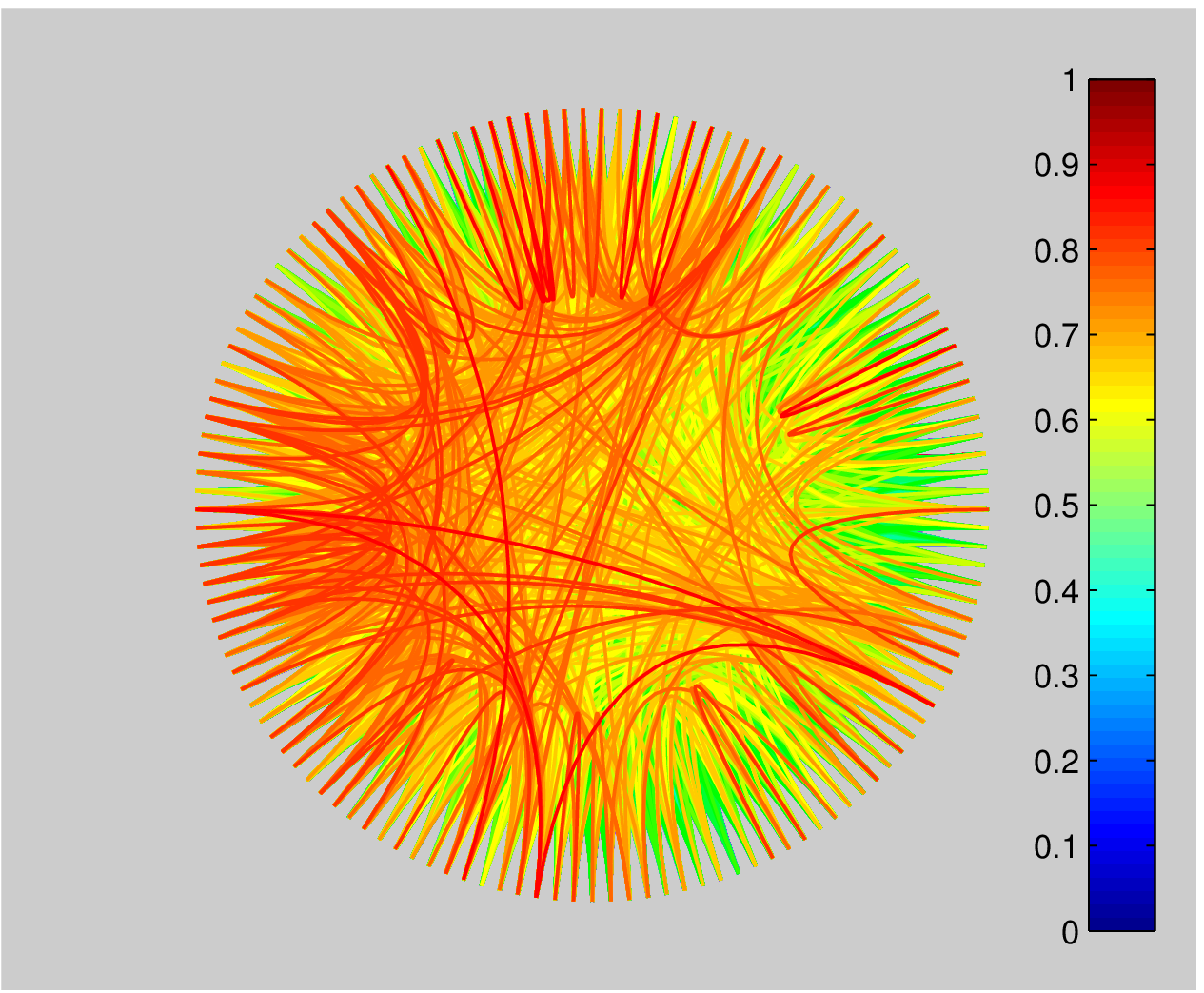}  
\caption{Shown are representative gene expression networks for a) complete remission,
b) refractory disease, and c) relapse. The nodes are the 133 cancer related genes and 
the links are calculated from Eq.\ (1) from one group of 16 patients from each of the three
patient categories.  The CCC values are $CCC_{\rm remission} = 0.2270
< CCC_{\rm refractory} = 0.2601
< CCC_{\rm relapse} = 0.5140 $.
\label{example}
}
\end{center} 

\end{figure}

In order to make comparisons between these groups that contain a different number of patients each, we chose 16 patients randomly from each group and constructed the gene network. This random selection of patients from each category helps to mitigate the bias due to different group sizes. We repeated this procedure 100 times which gave us 100 networks for each group. Error bars were calculated using this bootstrap procedure.  Representative networks are shown in Figure \ref{example}.

\subsubsection{Low, intermediate, and high risk groups}
The karyotype was used to classify AML patients 
into low risk, intermediate risk, or high risk categories depending on the cytogenetic aberration reported for each patient \cite{Bullinger2004}. We thus had three categories of patients: low risk group consisting of 38 patients, intermediate risk group consisting of 57 patients, and high risk group consisting of 16 patients. To make comparisons between these groups that contain a different number of patients each, we randomly chose 12 patients from each category and constructed the cancer related gene network. We repeated this procedure 100 times to mitigate any bias due to unequal sample sizes. This gave us 100 networks for each risk category. We calculated error bars using this bootstrap procedure.

\subsubsection{Groups with cancer at different levels of progression}
68 of the 116 patients died during the follow up period. The overall survival since diagnosis had been reported for each of these patients \cite{Bullinger2004}. We sorted the survival time dataset in descending order and divided it into 6 groups, each consisting of 16.67\% of the dataset. 
Samples were assigned into
groups (sextiles) based on length of survival from initial cancer diagnosis.
We constructed the gene network for each of these groups as stated above using Eq.\ (1) to define links between nodes.

\subsection{Gene network for each patient}
We constructed a network with 133 cancer related genes as nodes for each patient in our dataset. Two different methods were used to define the weights of the links in the network.

\subsubsection{Deviation of gene expression from average expression profile of AML cells}
We defined the weight of the link between any two nodes $i$ and $j$ in the network for patient $\alpha$ as follows:
\begin{equation}
l_{i,j}^\alpha=\sum_{k=1}^{116} \vert X_{k}^i-X_{k}^j \vert
\label{2}
\end{equation}
where $X_k^i =$ (expression of gene $i$ in patient $k$ - expression of
gene $i$ in patient $\alpha$)$^2$, i.e. the deviation of gene expression 
$i$ for patient $\alpha$ from the average
in the entire cancer dataset \cite{Bullinger2004}. 
We will show that the structure of such a network can discriminate between the patients that have cancer relapse and those that do not.

\subsubsection{Deviation of gene expression from average expression in different normal human tissues}
We used the gene expression data previously collected from 79 normal human tissues \cite{Su2004}. Cancer cells exhibit a gene expression profile different from cells in normal human tissues due to various mutations. We were motivated to study the deviations from normal expression profiles in different patients and the level of organization in these deviations. We, therefore, constructed a network with 133 cancer related genes as nodes and defined the weight of a link between nodes $i$ and $j$ as follows:
\begin{equation}
l_{i,j}^\alpha=\sum_{k=1}^{79} X_{k}^i *X_{k}^j
\label{3}
\end{equation}
where $X_k^i =$ 
(average expression of gene $i$ in normal tissue $k$ - expression of gene $i$ in cancer
cells in patient $\alpha$)$^2$, i.e.\ 
the deviation of expression of gene $i$ in patient $\alpha$ from the average expression of gene $i$ in normal human tissues \cite{Su2004}. 

\subsection{The $CCC$}
To quantify the structure in these gene networks, we defined a measure of hierarchy that quantifies how ``tree-like'' a network is \cite{Man}. The motivation behind this is that a tree network topology is the typical hierarchical structure. To calculate this measure, we computed the distance matrix defined by the network. The distance between nodes $i$ and $j$, $d_{i,j}$, is defined by the square root of the commute time. The commute time is the expected time taken by a random walker to travel from one of the nodes to the other and back to the starting node \cite{Saerens2004}.  Besides the weight of the link between the nodes $i$ and $j$, the commute time between nodes $i$ and $j$ also depends on all possible paths between the two nodes in the graph. In a weighted graph, the commute time between two nodes decreases with an increase in the number of possible paths from one node to the other and increases with an increase in the length of any path connecting the two nodes. Due 
 to these properties, the commute time is well suited for clustering tasks. To define the commute time, we let $L$ denote the graph Laplacian, defined as $L = D - A$, 
where $A$ is the matrix of links, $A = l$ 
in Eq.\ (1), (2) or (3), and the diagonal matrix
$D = {\rm diag} (A_i)$, with $A_i =\sum_j A_{ij}$.
 The commute time is obtained using $L_+$, the Moore-Penrose pseudoinverse of the graph Laplacian $L$ by   \cite{Barnett1990}
\begin{equation}
n(i, j)=V_G(e_i -e_j )^TL_+(e_i -e_j ).
\end{equation}
Here $(e_i)_j =\delta_{ij}$, and $V_G =\sum_{ij} a_{ij} $. Since  $L_+$ is symmetric and positive semidefinite,
$d_{ij} =\sqrt{n(i, j)}$ is a Euclidean distance metric, called the Euclidean commute time (ECT) distance.

To construct a tree topology that best approximates our original network, we applied the average linkage hierarchical clustering algorithm  \cite{Sokal}. This method takes the distance matrix defined by the gene network and produces a tree topology that best reproduces the original distances. The tree topology has the same nodes as the original network but different weights for the links between nodes. We calculated the correlation between the distances between all pairs of nodes in the original network and the distances between these pairs in the best-fitting tree. This gives the cophenetic correlation coefficient ($CCC$). The higher the correlation, the better does the tree topology approximate the original network and, thus, the more hierarchical is the original network. The best fitting tree to the distance metric data defines an approximation to the original network, termed the cophenetic matrix. The element $c_{ij}$ of the cophenetic matrix is the height where the network nodes $i$ and $j$ become members of the same cluster in the tree. It defines the distance between the nodes $i$ and $j$. The $CCC$ is defined as
\begin{equation}
CCC= \frac{\sum_{i<j}(d_{ij}-d)(c_{ij}-c)}{\sqrt{(\sum_{i<j}(d_{ij}-d)^2\sum_{i<j}(c_{ij}-c)^2}}
\label{ccc}
\end{equation}
Here $d$ is the average of the distances in the original network, $d_{ij}$, and $c$ is the average of the tree distances, $c_{ij}$.

\section{Results}
\begin{figure}[tbh!]  
\begin{center}  
a)\includegraphics[scale=0.35]{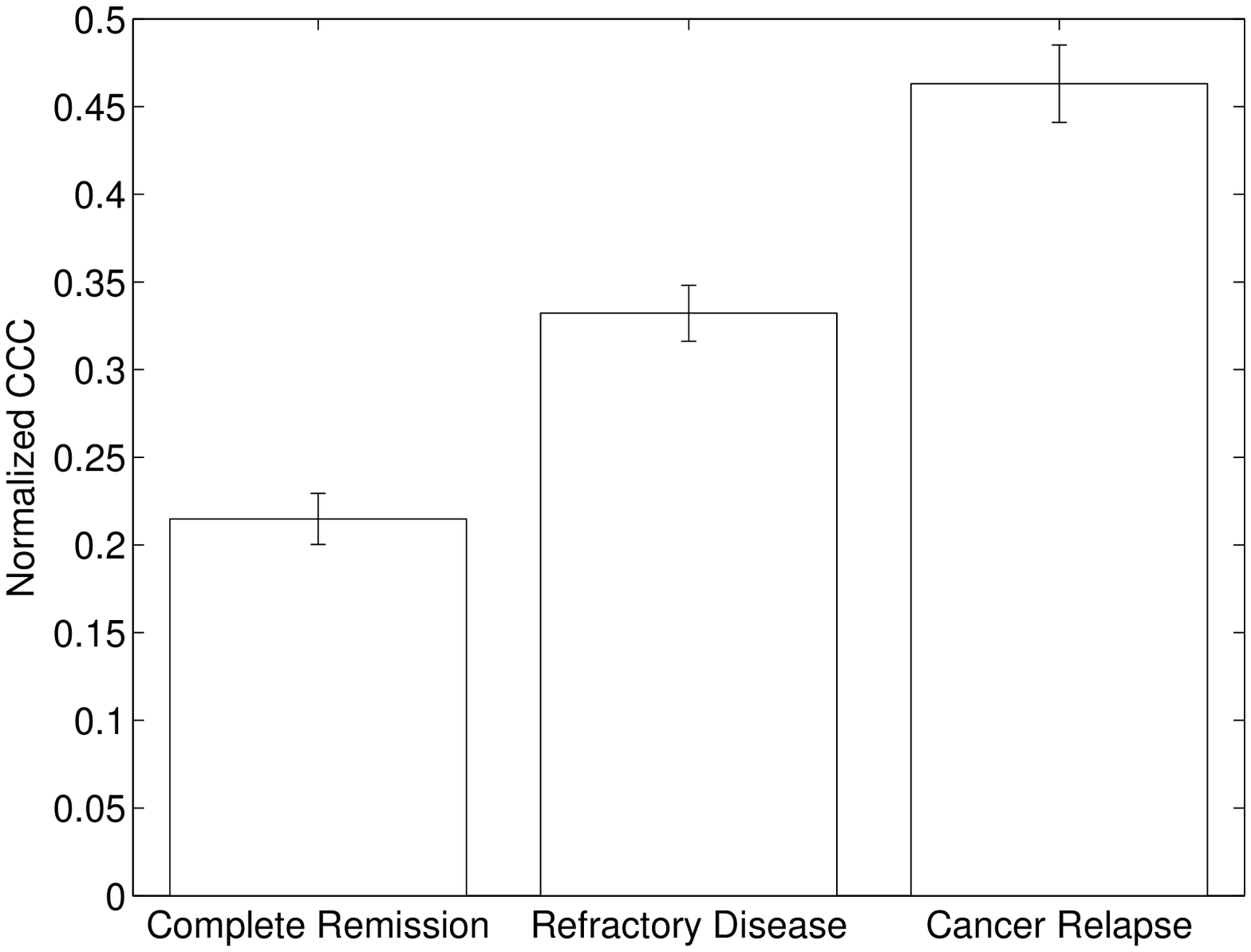}  
b)\includegraphics[scale=0.35]{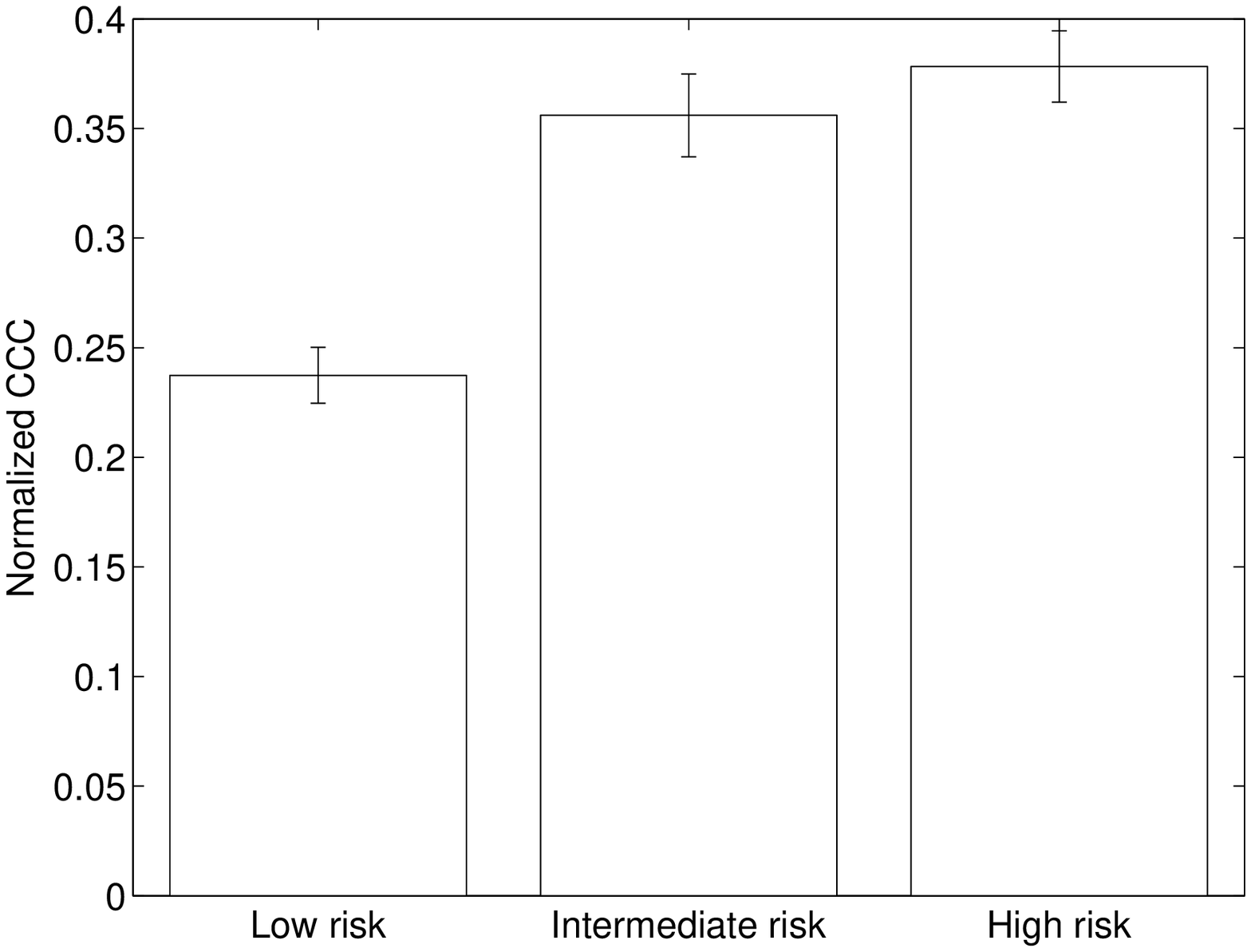}
\caption{a) The mean normalized $CCC$ of the cancer related gene network for the three categories of patients: patients that have a complete remission with no relapse during the follow up, patients that have a refractory disease, and patients that have a cancer relapse during the follow up. The links between nodes were calculated using Eq.\ (1).
b) The mean normalized $CCC$ of the cancer related gene network for different risk groups: low risk, intermediate risk, and high risk. The links between nodes were calculated using Eq.\ (1).  Error bars are one standard error.
\label{fig1}}
\end{center} 

\end{figure}

\subsection{Clinical outcome groups}
\begin{figure}[tbh!]   
\begin{center}  
\includegraphics[scale=0.45]{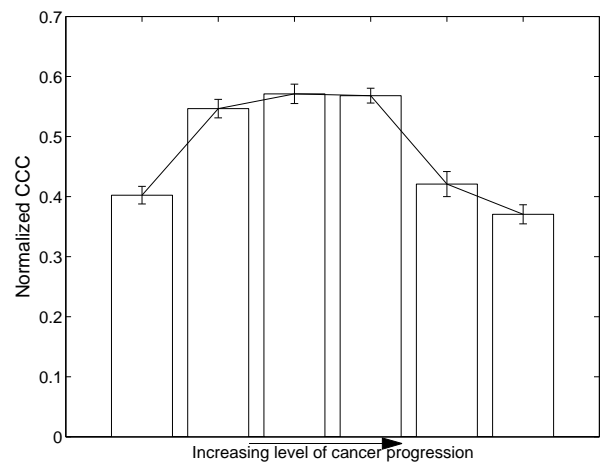}  

\caption{The mean normalized $CCC$ in the cancer related gene network for patients in different sextiles of survival time from initial diagnosis. The hierarchy in the network increases, peaking in the intermediate sextiles of survival time.
\label{fig2}}
\end{center} 

\end{figure}
We constructed networks comprising of 133 cancer related genes as nodes and the links calculated using Eq.\ (1) for each out of the three outcome categories: relapse, refractory disease, and complete remission with no relapse during follow up. Figure \ref{fig1}a shows the results. We compared these $CCC$ values with those from a random network with the same number of nodes and edges as the cancer related gene networks. We built 100 such random networks with redistributed link values. We defined a normalized cophenetic correlation coefficient as follows: $CCC_{norm} = (CCC-CCC_{rand})/(1-CCC_{rand})$ where $CCC_{rand}$ is the average $CCC$ for a random network of the same size. Normalized $CCC$ values are shown in Figure \ref{fig1}a. We computed the $z$-scores of the cancer-associated network $CCC$ relative to the distribution of the $CCC$ values of the random networks of the same size and sparsity: $Z_{CCC}=(CCC-CCC_{rand})/\sigma_{rand}$ . The $z$-scores for remission (no relapse), refractory disease, and relapse are $1.76, 2.84$, and $4.05$, respectively. The difference between remission and relapse groups is significant for a Student's t-test with a $p$-value $< 0.001$.

\subsection{Risk groups}
Figure \ref{fig1}b shows the normalized $CCC$ values for the three categories calculated as before. The high risk category has the maximum hierarchy among the three groups followed by intermediate, and low risk groups. The $z$-scores for low, intermediate, and high risk groups are $2.45,3.67$, and $3.90$, respectively.

\subsection{Cancer Progression}

We again constructed the cancer related gene network for patients exhibiting different sextiles of survival time and calculated the average normalized CCC and standard error in this average. The results are as shown in Figure \ref{fig2} wherein the independent axis indicates the time period between AML diagnosis and death, which we interpret as a measure of cancer progression, and the dependent axis indicates the mean normalized CCC. We observe that the variation of mean $CCC$ with increasing levels of cancer progression is non-monotonic. This result
makes manifest the non-trivial changes that occur in the cancer related gene 
expression network as a function of remaining survival time.

\subsection{Gene network for individual patients}

We constructed a cancer related gene network for each of the 116 patients with 133 cancer related genes as nodes and links defined in two different ways:

\subsubsection{Deviation of cancer cells gene expression from average expression profile of AML cells}
\begin{figure}[tbh!]   
\begin{center} 
\includegraphics[scale=0.45]{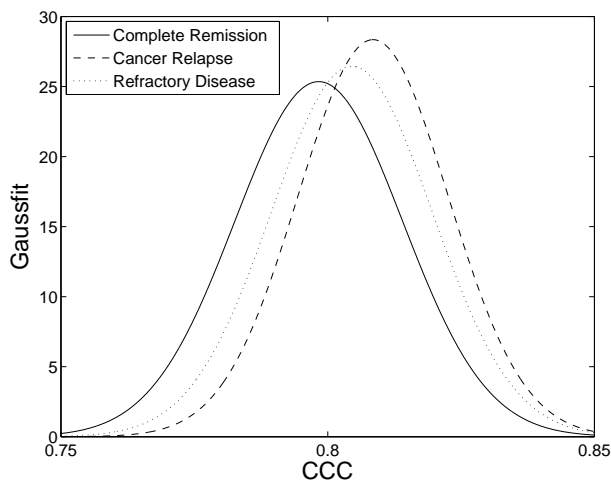}  

\caption{The Gaussian distribution fit to $CCC$ values for the patients in different clinical outcome categories: patients that have a complete remission with no relapse during the follow up (solid), patients that have a refractory disease (dotted), and patients that have a cancer relapse during the follow up (dashed). We constructed a cancer related gene network for each patient in the dataset with 133 cancer related genes as nodes and links calculated using Eq.\ (1).
\label{fig3}}
\end{center} 

\end{figure}

The links between nodes were calculated using Eq.\ (2). The distribution of $CCC$ values for relapse, remission, and refractory disease are shown in Figure \ref{fig3}. The difference between patients exhibiting relapse and those exhibiting complete remission is statistically significant for a Student's t-test with a $p$-value of $0.0019$. We plot a $ROC$ curve for the prediction that patients with $CCC > $ cutoff have a relapse. The area under the $ROC$ curve $(AUC)$ is $0.67$ which is significant compared to a random classifier.

\subsubsection{Deviation of cancer cells gene expression from average expression in different healthy human tissues}
\begin{figure}[tbh!]   
\begin{center}  
a)\includegraphics[scale=0.35]{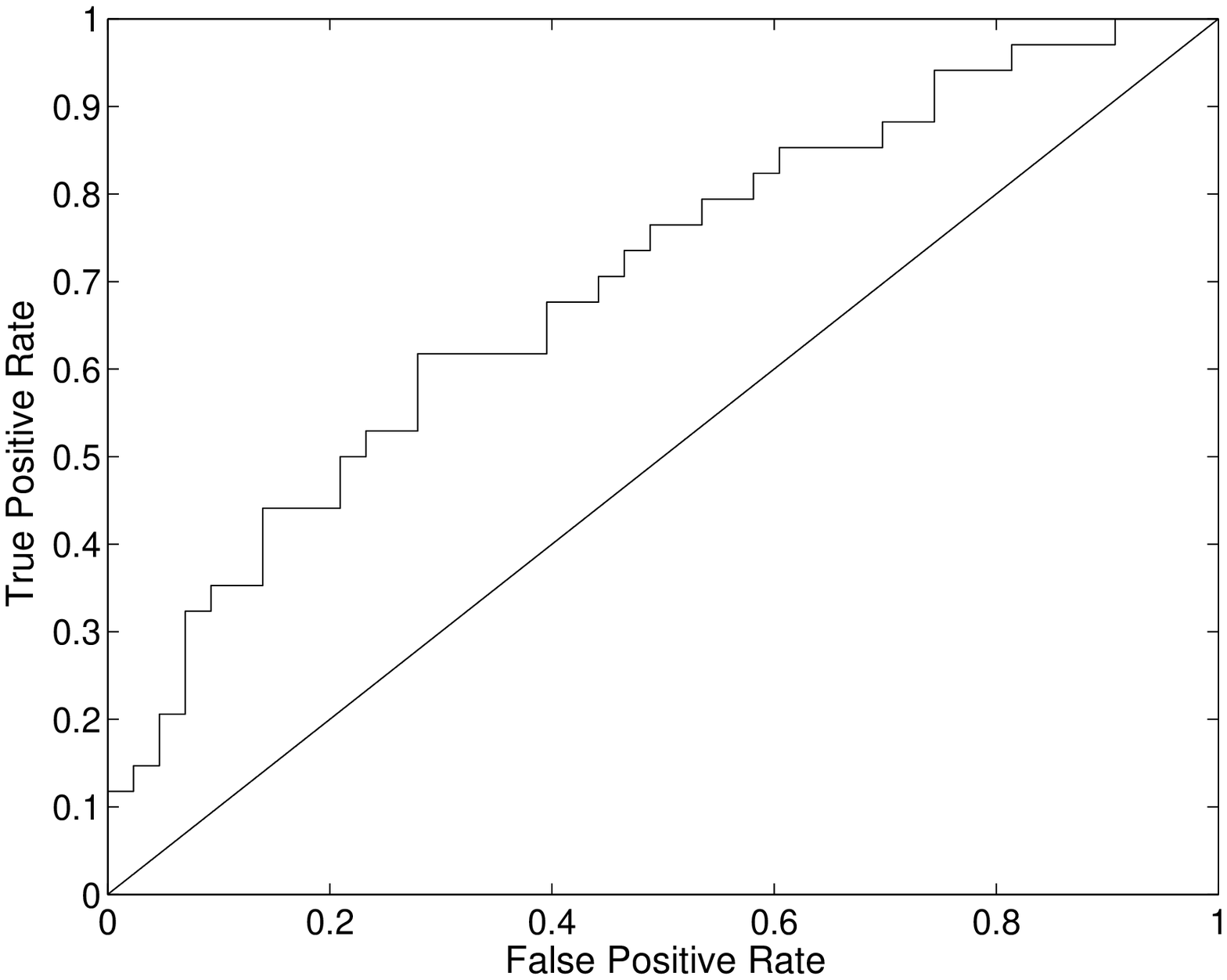}  
b)\includegraphics[scale=0.35]{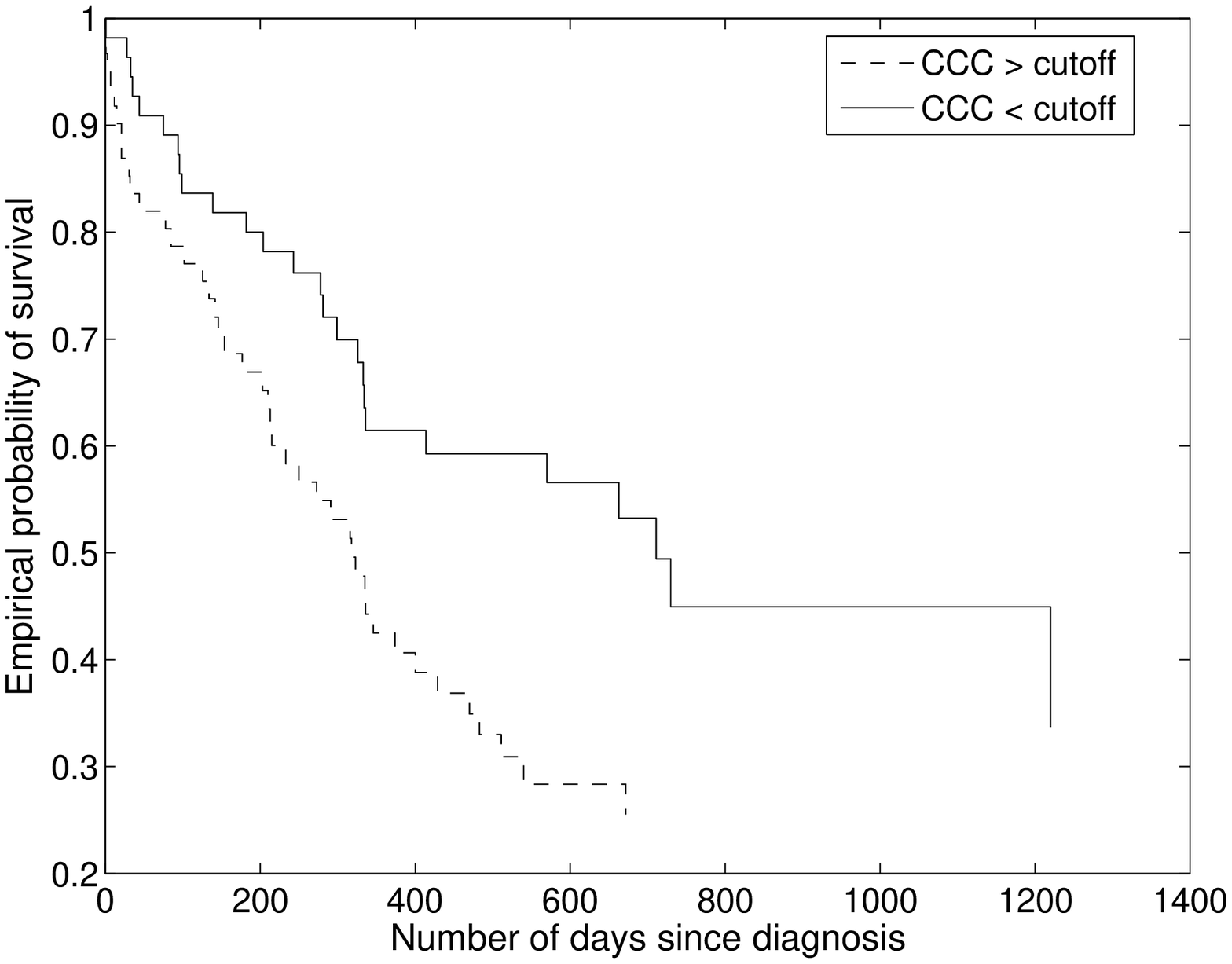}
\caption{a) The receiver operating characteristic (ROC) curve for the prediction that $CCC >$ cutoff leads to future relapse. The $CCC$ is calculated for the cancer related gene network with links calculated using Eq.\ (2). The area under the curve$(AUC)$ is $0.70$ which is significant as compared to a random classifier. b) The probability of survival for patients with $CCC > $ cutoff (dashed) and patients with $CCC < $ cutoff (solid). For patients alive at the end of follow up ($n=48$), survival time represents the number of days to last follow up. The hazard ratio is $2.04$. Thus, higher $CCC$ implies lower chances of survival as compared to patients with lower $CCC$ in the cancer related gene network. Here, cutoff = mean $CCC$ for all 116 patients. 
\label{fig4}}
\end{center} 

\end{figure}

The links between nodes are calculated using Eq.\ (3). The difference between patients exhibiting relapse and those exhibiting complete remission is significant with a $p$-value of $0.0012$. We plot a $ROC$ curve for the prediction that patients with $CCC > $ cutoff have a relapse, shown in Figure \ref{fig4}a. The area under the $ROC$ curve $(AUC)$ is $0.70$ which is significant compared to a random classifier.

We further used the $CCC$ values for individual patient gene networks to predict the risk of death within 3 years of cancer diagnosis. For this purpose, we used the cancer related gene network for each patient with links calculated using Eq.\ (2). We calculated the $CCC$ values for 68 patients who died during the follow up period and for the patients who were alive at the end of follow up. The difference between the two categories is statistically significant for a Student's t-test with a $p$-value $< 0.001$. Using these $CCC$ values for the two groups, we attempted to predict the chances of death within 3 years of cancer diagnosis by setting a cutoff on the $CCC$ values. The $AUC$ of the $ROC$ curve for this prediction is $0.69$ which is again significant as compared to a random classifier.


\section{Discussion}

There are selection pressures on a tumor that
are distinct from those upon the host.
The proliferation and progression of cancer involves activation of 
cancer-specific gene pathways, perhaps hijacked ancestral host 
pathways \cite{Davies2011}. 
Thus, in contrast to the classical view of cancer as a dedifferentiation
of the host, progression of a tumor may involve structuring of specific
expression pathways.
For example, we have here suggested that a more aggressive and resistant
 cancer is expected to have a more structurally organized and modular cancer-associated network. This is, indeed, observed in Figure \ref{fig1}a. The average $CCC$ is the highest for patients that exhibit a cancer relapse during the follow up. Post remission, there is competition between the residual cancerous cells and the normal or transplanted stem cells in the bone marrow. To compete against and to drive out these cells, there is selection for a more organized and modular cancer related genes network among the AML cells. Hence, we expected higher hierarchy in the gene network for patients that exhibit cancer relapse during follow up. For patients that exhibit complete cancer remission and no relapse, the average $CCC$ is the lowest. In these patients, the cancer cells are completely eradicated during chemotherapy. Any residual cells in the bone marrow, if present, are unable to compete against the normal stem cells and are unable to start proliferating again aggressively enough for cancer to relapse. Therefore, we expected cancer cells in these patients to exhibit the lowest level of organization in the cancer related genes network. In some AML cases, the cancer cells evolve mechanisms to resist the entry of anti-cancer chemo drugs into the cells. These cancers are thus resistant to existing chemotherapy procedures, and such patients are said to have a refractory cancer disease. In such cases, there is less selection pressure on AML cells since these are already resistant to anti-cancer drugs. Hence, we expected the cancer related gene network in such patients to be less organized as compared to patients that exhibit relapse. On the other hand, since these cancer cells are drug resistant and also compete with the normal bone marrow cells, they exhibit a higher level of organization as compared to cancer cells from patients exhibiting complete remission.

Cancers exhibiting a higher level of organizational hierarchy and modularity in the cancer related genes network are expected to be more competitive, proliferative, and aggressive and, thus, present a higher level of risk for the host. This is, indeed, observed in Figure \ref{fig1}b wherein the average $CCC$ increases with increasing risk. Here, the different risk categories were determined depending of the cytogenetic profile of the patients \cite{Grimwade1998,Byrd2002,Slovak2000,Bullinger2004}.

We also calculated the average $CCC$ of the cancer related gene network for patients as a function of survival time since initial cancer diagnosis. The results are shown in Figure \ref{fig2}. As cancer progresses, cancer cells compete with the host cells for oxygen and nutrition. The host immune system attempts to resist or eradicate tumor progression \cite{frey,finn, vesley, suzuki}. Cancer cells thus evolve in a rugged fitness landscape. Therefore, there is selection for more effective, aggressive, and proliferative gene pathways since cancer cells evolve amidst tremendous environmental pressure from the host cells and the host immune system \cite{vito}. During evolution in such a rugged fitness landscape, theory shows that modularity in gene networks is selected for \cite{Sun2007,Deem}. Therefore, the cancer related gene network becomes more organized and modular as cancer progresses and matures. However, once the cancer  reaches a mature phase, the host immune system is severely compromised, especially in the cancer microenvironment \cite{dieu,buggins}. Also, cancer cells now dominate the tissue environment with their high proliferation rates, induced angiogenesis and reprogrammed energy metabolism \cite{hanahan}. Therefore, there is less selection pressure since the cancer cells do not have to compete as much with the host marrow cells or the impaired and compromised immune system. This presents a relatively smooth fitness landscape for cancer cell evolution with little or no selection, and theory shows that modularity is selected for less strongly \cite{Sun2007,Deem}.  Thus, we expect the level of organization of the cancer associated gene network to decline
once the cancer has dominated the host.
That is, we suggest that the level of organization in the cancer related gene networks declines due to less selection pressure requiring this costly structure in the advanced phase of the cancer. 
Modifications to the expression network of the cancer associated genes may
stem from mutations, compromised DNA replication, or
other mechanisms in cancer cells \cite{negrini}.  
We suggest that nonmonotonic selection pressure may
be the reason for the nonmonotonic variation of modularity with cancer progression seen in Figure \ref{fig2}.

We also attempted to differentiate between patients that exhibit relapse during follow up from patients that do not show any relapse using the level of organization in cancer related gene network in their cancerous cells at the time of diagnosis. For this, we constructed a cancer related gene network for each patient with 133 cancer related genes as nodes and links defined with either Eq. (2) or (3). We then set a cutoff on the $CCC$ values to discriminate between patients with relapse and non-relapse cases. The area under the $ROC$ curve is $0.69$ and $0.70$, respectively, for the two cases which is significant as compared to a random classifier. We also constructed a cancer related gene network for each patient using Eq.\ (3), considering deviation of gene expression in cancer cells from normal, non-cancerous, healthy bone marrow, and myeloid blood cells only. When classifying using $CCC$ values calculated from these networks, the area under the $ROC$ curve (not shown) was $0.65$. Similarly, we attempted to predict the chances of death for a patient within 3 years of diagnosis using the $CCC$ value of the cancer related gene network at the time of diagnosis. We again set a cutoff on the $CCC$ values to discriminate between patients who died during the follow up from patients that were alive at the end of follow up. The $AUC$ is $0.69$ which is again statistically significant in comparison with a random classifier. The $CCC$ of the cancer related gene network in the cancerous cells of a patient is a biomarker in AML prognosis. It is discriminating and may help identify patients that are at a higher risk of a relapse after a typical chemotherapy regime before the treatment has begun. This biomarker may also help to identify the overall risk that the patient is subject to, which may be helpful in deciding upon a chemotherapy regime for each patient.
 
\section{Conclusion}
We defined a measure of hierarchical organization in gene pathways in AML cancer cells. 
With a retrospective cohort analysis based on the gene expression profiles of 116 acute myeloid leukemia patients,
we found that the level of organization, indicated by $CCC$ values, correlates with the clinical outcome in AML and with the level of risk the cancer presents for the host. It also correlates with the survival period in the case of dead patients. We expected this result since a more organized and modular network of cancer associated genes is likely to contribute towards a more aggressive form of cancer. Further, since cancer cells evolve in a rugged fitness landscape, a more organized cancer related gene network will be selected for amidst the immense selection pressure in a changing, hostile environment  \cite{Sun2007,Deem,Albert}. We also showed how the level of organization in cancer associated gene pathways varies as cancer progresses and matures. This variation is non-monotonic, likely due to a decrease in the selection pressure once the cancer dominates. Finally, the $CCC$ calculated for each patient can serve as a biomarker in AML prognosis to assess the risk of relapse and overall risk level for a patient and thus to inform a suitable personalized treatment regime.

\section*{Acknowledgment}
This work was supported by the US National Institutes of Health under grant 1 R01 GM 100468-01.
Shubham Tripathi is a recipient of the Khorana Fellowship for the year 2014.


\section*{Supplemental Information}
Files listing the 133 cancer-related genes studied as well as 
clinical outcome, risk group,  and CCC for each patient
are available as supplemental information.

\bibliography{aml}

\end{document}